\documentclass[journal]{IEEEtran}
\usepackage[font={footnotesize}]{caption}
\usepackage{subcaption}
\usepackage{cite}
\usepackage{balance}
\usepackage{amsmath,amssymb,amsfonts}
\usepackage{algorithmic}
\usepackage{graphicx}
\usepackage{graphicx}
\usepackage{textcomp}
\usepackage{xcolor}
\usepackage{tikz}
\usepackage{pgfplots} 
\usepackage{tikzscale}
\usepackage{pgfplots}
\usepackage{cite}
\usepackage{amsmath,amssymb,amsfonts}
\usepackage{algorithmic}
\usepackage{textcomp}
\usepackage{accents} 
\usepackage{xcolor}

\usepackage{acro}

\DeclareAcronym{5g}{
short=5G,
long= fifth generation,
}
\DeclareAcronym{6g}{
short=6G,
long= sixth generation,
}
\DeclareAcronym{3d}{
short=3D,
long= three-dimensional,
}
\DeclareAcronym{aod}{
short=AOD,
long= angle-of-departure,
}
\DeclareAcronym{aosa}{
short=AOSA,
long= array-of-subarray,
}
\DeclareAcronym{adod}{
short=ADOD,
long= angle-difference-of-departure,
}
\DeclareAcronym{aoa}{
short=AOA,
long= angle-of-arrival,
}
\DeclareAcronym{adc}{
short=ADC,
long= analog to digital converter,
}
\DeclareAcronym{aeb}{
short=AEB,
long= angle error bound,
}
\DeclareAcronym{av}{
short=AV,
long= autonomous vehicle,
}
\DeclareAcronym{bs}{
short=BS,
long= base station,
}
\DeclareAcronym{bse}{
short=BSE,
long= beam split effect,
}
\DeclareAcronym{csi}{
short=CSI,
long= channel state information,
}
\DeclareAcronym{cfo}{
short=CFO,
long= carrier frequency offset,
}
\DeclareAcronym{ceb}{
short=CEB,
long= clock error bound,
}
\DeclareAcronym{coa}{
short=COA,
long= curvature-of-arrival,
}
\DeclareAcronym{crb}{
short=CRB,
long= Cram\'er-Rao bound,
}
\DeclareAcronym{ccrb}{
short=CCRB,
long= constrained Cram\'er-Rao bound,
}
\DeclareAcronym{cmos}{
short=CMOS,
long= complementary metal-oxide-semiconductor,
}
\DeclareAcronym{crlb}{
short=CRLB,
long= Cram\'er-Rao lower bound,
}
\DeclareAcronym{cdf}{
short=CDF,
long= cumulative distribution function,
}
\DeclareAcronym{cp}{
short=CP,
long= cyclic prefix,
}
\DeclareAcronym{dac}{
short=DAC,
long= digital to analog converter,
}
\DeclareAcronym{dfl}{
short=DFL,
long= device-free localization,
}
\DeclareAcronym{dmimo}{
short=D-MIMO,
long= distributed MIMO,
}
\DeclareAcronym{dlprs}{
short=DL-PRS,
long= downlink positioning reference signal,
}
\DeclareAcronym{d2d}{
short=D2D,
long= device-to-device,
}
\DeclareAcronym{dftsofdm}{
short=DFT-s-OFDM,
long= discrete-Fourier-transform spread OFDM,
}
\DeclareAcronym{dl}{
short=DL,
long= deep learning,
}
\DeclareAcronym{em}{
short=EM,
long= electromagnetic,
}
\DeclareAcronym{gps}{
short=GPS,
long= global positioning system,
}
\DeclareAcronym{hwi}{
short=HWI,
long= hardware impairment,
}
\DeclareAcronym{hemt}{
short=HEMT,
long= high electron mobility transistor,
}
\DeclareAcronym{hbt}{
short=HBT,
long= heterojunction bipolar transistors,
}
\DeclareAcronym{iot}{
short=IoT,
long= internet of things,
}
\DeclareAcronym{isac}{
short=ISAC,
long= integrated sensing and communication,
}
\DeclareAcronym{iqi}{
short=IQI,
long= in-phase and quadrature imbalance,
}
\DeclareAcronym{ia}{
short=IA,
long= initial access,
}
\DeclareAcronym{kpi}{
short=KPI,
long= key performance indicator,
}
\DeclareAcronym{kf}{
short=KF,
long= Kalman filter,
}
\DeclareAcronym{ekf}{
short=EKF,
long= extended Kalman filter,
}
\DeclareAcronym{ukf}{
short=UKF,
long= unscented Kalman filter,
}
\DeclareAcronym{ckf}{
short=CKF,
long= cubature Kalman filter,
}
\DeclareAcronym{pf}{
short=PF,
long= particle filter,
}
\DeclareAcronym{lb}{
short=LB,
long= lower bound,
}
\DeclareAcronym{lse}{
short=LSE,
long= least-square estimator,
}
\DeclareAcronym{lo}{
short=LO,
long= local oscillator,
}
\DeclareAcronym{mc}{
short=MC,
long= mutual coupling,
}
\DeclareAcronym{mac}{
short=MAC,
long= medium access control,
}
\DeclareAcronym{meb}{
short=MEB,
long= mapping error bound,
}
\DeclareAcronym{ml}{
short=ML,
long= machine learning,
}
\DeclareAcronym{mcrb}{
short=MCRB,
long= misspecified Cram\'er-Rao bound,
}
\DeclareAcronym{mds}{
short=MDS,
long= multidimensional scaling ,
}
\DeclareAcronym{mimo}{
short=MIMO,
long= multiple-input-multiple-output,
}
\DeclareAcronym{mm}{
short=MM,
long= mismatched model,
}
\DeclareAcronym{mpc}{
short=MPC,
long= multipath components,
}
\DeclareAcronym{mmwave}{
short=mmWave,
long= millimeter wave,
}
\DeclareAcronym{mmle}{
short=MMLE,
long= mismatched maximum likelihood estimation,
}
\DeclareAcronym{mems}{
short=MEMS,
long= micro-electro-mechanical system,
}
\DeclareAcronym{mle}{
short=MLE,
long= maximum likelihood estimation,
}
\DeclareAcronym{nlos}{
short=NLOS,
long= none-line-of-sight,
}
\DeclareAcronym{ofdm}{
short=OFDM,
long= orthogonal frequency-division multiplexing,
}
\DeclareAcronym{oeb}{
short=OEB,
long= orientation error bound,
}
\DeclareAcronym{otfs}{
short=OTFS,
long= orthogonal time-frequency space,
}
\DeclareAcronym{pdf}{
short=PDF,
long= probability density function,
}
\DeclareAcronym{papr}{
short=PAPR,
long= peak-to-average-power ratio,
}
\DeclareAcronym{pan}{
short=PAN,
long= power amplifier nonlinearity,
}
\DeclareAcronym{pa}{
short=PA,
long= power amplifier,
}
\DeclareAcronym{ps}{
short=PS,
long= phase shifter,
}
\DeclareAcronym{pn}{
short=PN,
long= phase noise,
}
\DeclareAcronym{poa}{
short=POA,
long= phase-of-arrival,
}
\DeclareAcronym{pwm}{
short=PWM,
long= planar wave model,
}
\DeclareAcronym{pdoa}{
short=PDOA,
long= phase-difference-of-arrival,
}
\DeclareAcronym{prs}{
short=PRS,
long= positioning reference signals,
}
\DeclareAcronym{peb}{
short=PEB,
long= position error bound,
}
\DeclareAcronym{rnn}{
short=RNN,
long= recurrent neural network,
}
\DeclareAcronym{rl}{
short=RL,
long= reinforcement learning,
}
\DeclareAcronym{rf}{
short=RFC,
long= radio-frequency,
}
\DeclareAcronym{rfid}{
short=RFID,
long= radio frequency identification,
}
\DeclareAcronym{ris}{
short=RIS,
long= reconfigurable intelligent surface,
}
\DeclareAcronym{rss}{
short=RSS,
long= received signal strength,
}
\DeclareAcronym{rtt}{
short=RTT,
long= round-trip time,
}
\DeclareAcronym{sm}{
short=SM,
long= standard model,
}
\DeclareAcronym{sige}{
short=SiGe,
long= silicon-germanium,
}
\DeclareAcronym{spp}{
short=SPP,
long= surface plasmon polariton,
}
\DeclareAcronym{sa}{
short=SA,
long= subarray,
}
\DeclareAcronym{sota}{
short=SOTA,
long= state-of-the-art,
}
\DeclareAcronym{swm}{
short=SWM,
long= spherical wave model,
}
\DeclareAcronym{slam}{
short=SLAM,
long= simultaneous localization and mapping,
}
\DeclareAcronym{tm}{
short=TM,
long= true model,
}
\DeclareAcronym{toa}{
short=TOA,
long= time-of-arrival,
}
\DeclareAcronym{tof}{
short=TOF,
long= time-of-flight,
}
\DeclareAcronym{tdoa}{
short=TDOA,
long= time-difference-of-arrival,
}
\DeclareAcronym{thz}{
short=THz,
long= terahertz,
}
\DeclareAcronym{ue}{
short=UE,
long= user equipment,
}
\DeclareAcronym{ummimo}{
short=UM-MIMO,
long= ultra-massive multi-input-multi-output,
}
\DeclareAcronym{vlp}{
short=VLP,
long= visible light positioning,
}
\DeclareAcronym{veb}{
short=VEB,
long= velocity error bound,
}
\DeclareAcronym{vlc}{
short=VLC,
long= visible light communication,
}
\DeclareAcronym{ula}{
short=ULA,
long= uniform linear array,
}
\DeclareAcronym{upa}{
short=UPA,
long= uniform planar array,
}
\DeclareAcronym{wlan}{
short=WLAN,
long= wireless local area network,
}

\usepackage{pgfplots}
\usepackage{tikz}
\usetikzlibrary{calc}
\makeatletter
\newcommand{\gettikzxy}[3]{%
  \tikz@scan@one@point\pgfutil@firstofone#1\relax
  \edef#2{\the\pgf@x}%
  \edef#3{\the\pgf@y}%
}

\def\BibTeX{{\rm B\kern-.05em{\sc i\kern-.025em b}\kern-.08em
    T\kern-.1667em\lower.7ex\hbox{E}\kern-.125emX}}
\begin{document}

\title{
Joint User Localization and Location Calibration of A Hybrid Reconfigurable Intelligent Surface} 

\author{Reza~Ghazalian,~\IEEEmembership{Member,~IEEE,} 
Hui~Chen,~\IEEEmembership{Member,~IEEE,}
George~C~Alexandropoulos,~\IEEEmembership{Senior Member,~IEEE,}
Gonzalo~Seco-Granados,~\IEEEmembership{Senior Member,~IEEE,}
Henk~Wymeersch,~\IEEEmembership{Senior Member,~IEEE,} and~Riku~J{\"a}ntti,~\IEEEmembership{Senior~Member,~IEEE}
\thanks{This work has been funded in part by the Academy of Finland Profi-5 under the grant number 326346, ULTRA under grant number 328215,
and the EU H2020 RISE-6G project under grant number 10101701.}
\thanks{R. Ghazalian and R. Jäntti are with the Department of Communications and Networking, School of Electrical Engineering
of Electrical Engineering, Aalto University, 02150 Espoo, Finland  ( \{reza.ghazalian, riku.jantti\}@aalto.fi).}
\thanks{H. Chen and H. Wymeersch are with the Department of Electrical Engineering, Chalmers University of Technology, 412 58 Gothenburg, Sweden (emails: \{hui.chen, henkw\}@chalmers.se).}
\thanks{G. C. Alexandropoulos is with the Department of Informatics and Telecommunications, National and Kapodistrian University of Athens, 15784 Athens, Greece (e-mail: alexandg@di.uoa.gr).}
\thanks{G. Seco-Granados is with the Department of Telecommunications and
Systems Engineering, Universitat Autònoma de Barcelona, 08193 Barcelona,
Spain (e-mail: gonzalo.seco@uab.cat).}
\vspace{-0.5cm}}

\maketitle
\begin{abstract}
The recent research in the emerging technology of reconfigurable intelligent surfaces (RISs) has identified its high potential for localization and sensing. However, to accurately localize a user placed in the area of influence of an RIS, the RIS location needs to be known a priori and its phase profile is required to be optimized for localization. In this paper, we study the problem of the joint localization of a hybrid RIS (HRIS) and a user, considering that the former is equipped with a single reception radio-frequency (RF) chain enabling simultaneous tunable reflections and sensing via power splitting. Focusing on the downlink of a multi-antenna base station, we present a multi-stage approach for the estimation of the HRIS position and orientation as well as the user position. Our simulation results, including comparisons with the Cramér-Rao lower bounds, demonstrate the efficiency of the proposed localization approach, while showcasing that there exists an optimal HRIS power splitting ratio for the desired multi-parameter estimation problem.

\end{abstract}


\section{Introduction}
\IEEEPARstart{R}{Econfigurable} intelligent surfaces (RISs) consist of multiple metamaterials of an almost passive nature \cite{huang2020holographic}, which can impact the characteristics of \ac{em} waves impinging on them, such as the phase, offering radio propagation control~\cite{WavePropTCCN}. The RIS-parametrized channel can assist communications via its potential beamforming gain and act as an additional passive anchor in improving localization performance~\cite{wymeersch2020radio}. Therefore, \acp{ris} are considered as a candidate technology for the sixth generation (6G) of wireless systems~\cite{RISE6G_COMMAG}, where joint localization and communication is expected to support various use cases~\cite{saad2019vision}, such as autonomous driving, digital twins, and other immersive applications.

The consideration of \acp{ris} for localization applications has received significant attention in recent years, 
see, e.g.,~\cite{abu2021near,huang2022near,keykhosravi2021siso}.
In \cite{abu2021near,huang2022near,keykhosravi2021siso}, two-dimensional (2D) and 3D user localization scenarios with either a far-field (FF) or near-field (NF) channel model, static or mobility conditions, as well as single-input-single-output or \ac{mimo} system system setups were considered showcasing the potential of RIS-assisted localization. Very recently, \cite{Keykhosravi2022infeasible_all} overviewed various scenarios where the adoption of RISs enables localization, which would be otherwise impossible. %

In the RIS-focused research, channel estimation is a challenging problem due to the cascaded channel structure between the \ac{bs} and the \ac{ue}, as well as the commonly considered passive nature of RISs, which deprives them from any estimation capability. Nevertheless, the various available cascade channel estimation techniques~\cite{Tsinghua_RIS_Tutorial} result in certain limitations on the wireless operation design \cite{alexandropoulos2021hybrid} (the lack of the individual channels complicates network management and the design of modulation and coding) and challenge wireless localization~\cite{wymeersch2020radio}. In addition, the vast majority of the RIS-based localization studies~\cite{abu2021near,huang2022near,keykhosravi2021siso,alexandropoulos2022localization} assume that the RIS state (mainly its position and orientation) is known, which can be nontrivial in various practical cases. The estimation of the RIS state was only recently addressed in~\cite{ghazalian2022bi}, however, the locations of both transmitters and receivers were assumed known, which requires certain overhead for infrastructure calibration. As a consequence, the high-accuracy joint location estimation of \acp{ue} and \acp{ris} is an open problem.

One of the main challenges with the latter estimation problem stems from the fact that the angle of arrival (AOA) and angle of departure (AOD) at the RIS are coupled and cannot be estimated separately. To tackle this issue, in this paper, we propose the adoption of an RIS equipped with a single reception (RX) \ac{rf} chain, instead of using a passive RIS. This RIS architecture, which was first presented in~\cite{alexandropoulos2020hardware} to enable the estimation of individual channels and was then extended in~\cite{alexandropoulos2021hybrid} to hybrid \acp{ris} capable of simultaneous reflection and sensing, considers wave-guides capable of feeding the impinging signals at each RIS unit element to base-band unit via a network of phase shifters which combines them to feed an RX \ac{rf} chain. We particularly consider an HRIS and focus on its joint localization with a UE in a 2D scenario under FF conditions. A  multi-stage estimator is presented whose performance is benchmarked by the Cramér-Rao lower bounds (CRBs). Our simulation results demonstrate the efficiency of the proposed estimation approach and showcase the impact of the HRIS power splitting ratio on the localization performance.
\subsubsection*{Notation}
Vectors and matrices are indicated by lowercase and uppercase bold letters, respectively. Notation $\left[\mathbf{A}\right]_{i,j}$ represents $\mathbf{A}$'s element in the $i$-th row and $j$-th column, while the index $i:j$ determines all the elements between $i$ and $j$. The complex conjugate, Hermitian, transpose, and Moore–Penrose inverse operators are represented by $\left( .\right)^*$, $\left( .\right)^{\mathsf{H}}$, $\left( .\right)^\top$, and $\left( .\right)^\dag$, respectively. $\Vert.\Vert$ returns the norm of vectors or the Frobenius norm of matrices, whereas $\odot$ and $\otimes$ indicate the element-wise and Kronecker products, respectively. $\mathbf{1}_K$ is a column vector comprising all ones with length $K$ and function $\text{atan2}(y,x)$ is the four-quadrant inverse tangent function.

\section{System and Channel Models}
\subsection{System Setup}
\begin{figure}
    \centering
    \begin{tikzpicture}
    \node (image) [anchor=south west]{\includegraphics[width=.95\columnwidth]{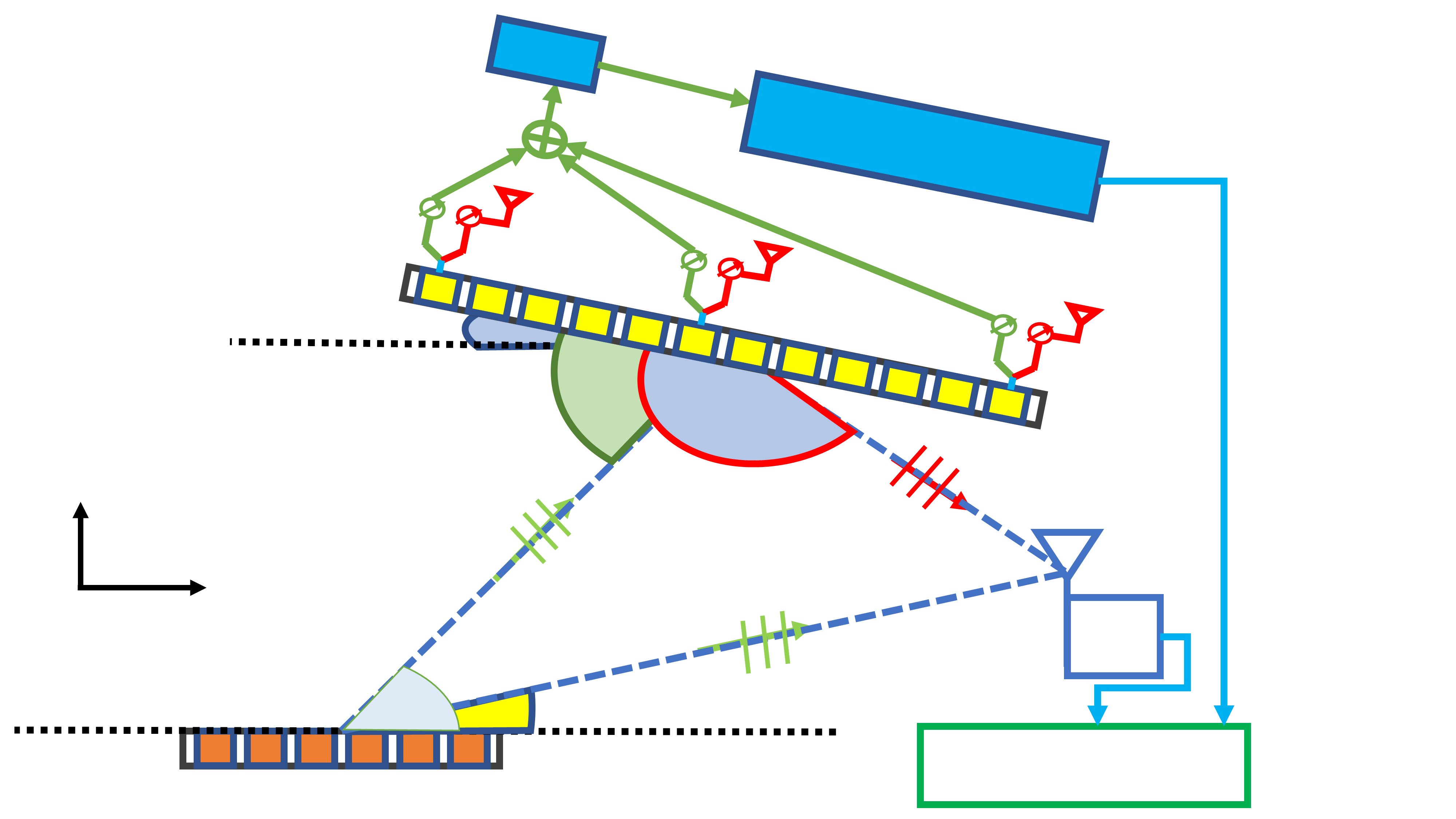}};
    \gettikzxy{(image.north east)}{\ix}{\iy};
    \node at (0.25*\ix,0.1*\iy)[rotate=0,anchor=north]{\footnotesize{BS}}; 
    \node at (0.33*\ix,0.26*\iy)[rotate=0,anchor=north]{\footnotesize{$\theta_{_\text{BR}}$}};
    \node at (0.41*\ix,0.21*\iy)[rotate=0,anchor=north]{\footnotesize{$\theta_{_\text{BU}}$}};
    \node at (0.49*\ix,0.56*\iy)[rotate=0,anchor=north]{\footnotesize{$\theta_{_\text{RU}}$}};
    \node at (0.41*\ix,0.59*\iy)[rotate=0,anchor=north]{\footnotesize{$\phi_{_\text{RB}}$}};
     \node at (0.30*\ix,0.64*\iy)[rotate=0,anchor=north]{\footnotesize{\shortstack{$\alpha$}}};
     \node at (0.26*\ix,0.34*\iy)[rotate=43,anchor=north]{\footnotesize{\shortstack{Incident signal}}};
     \node at (0.46*\ix,0.27*\iy)[rotate=15,anchor=north]{\footnotesize{\shortstack{LOS}}};
     \node at (0.75*\ix,0.13*\iy)[rotate=0,anchor=north]{\footnotesize{\shortstack{CPU}}};
     \node at (0.752*\ix,0.27*\iy)[rotate=0,anchor=north]{\footnotesize{\shortstack{UE}}};
     \node at (0.884*\ix,0.29*\iy)[rotate=0,anchor=north]{\footnotesize{\shortstack{Reliable \\ Link}}};
     \node at (0.68*\ix,0.46*\iy)[rotate=-40,anchor=north]{\footnotesize{\shortstack{NLOS}}};
      \node at (0.385*\ix,0.95*\iy)[rotate=-11,anchor=north]{\footnotesize{\shortstack{RFC}}};
      \node at (0.635*\ix,0.86*\iy)[rotate=-11,anchor=north]{\footnotesize{\shortstack{Digital Controller}}};
      \node at (0.843*\ix,0.69*\iy)[rotate=0,anchor=north]{\footnotesize{\shortstack{\textcolor{red}{Reflected Signal}}}};
      \node at (0.843*\ix,0.64*\iy)[rotate=0,anchor=north]{\footnotesize{\shortstack{\textcolor{red}{at Mth Element}}}};
      \node at (0.75*\ix,0.573*\iy)[rotate=-11,anchor=north]{\footnotesize{\shortstack{$1-\rho$}}};
      \node at (0.665*\ix,0.6*\iy)[rotate=-11,anchor=north]{\footnotesize{\shortstack{$\rho$}}};
      \node at (0.539*\ix,0.653*\iy)[rotate=-11,anchor=north]{\footnotesize{\shortstack{$1-\rho$}}};
      \node at (0.4540*\ix,0.6800*\iy)[rotate=-11,anchor=north]{\footnotesize{\shortstack{$\rho$}}};
      \node at (0.368*\ix,0.713*\iy)[rotate=-11,anchor=north]{\footnotesize{\shortstack{$1-\rho$}}};
      \node at (0.2830*\ix,0.74*\iy)[rotate=-11,anchor=north]{\footnotesize{\shortstack{$\rho$}}};
      \node at (0.42*\ix,0.72*\iy)[rotate=-11,anchor=north]{\footnotesize{\shortstack{$\dots$}}};
      \node at (0.62*\ix,0.65*\iy)[rotate=-11,anchor=north]{\footnotesize{\shortstack{$\dots$}}};
      \node at (0.15*\ix,0.3*\iy)[rotate=0,anchor=north]{\footnotesize{\shortstack{x}}};
      \node at (0.055*\ix,0.41*\iy)[rotate=0,anchor=north]{\footnotesize{\shortstack{y}}};
      \node at (0.55*\ix,.99*\iy)[rotate=-11,anchor=north]{\footnotesize{\shortstack{Sensed Signal at RFC}}};
    \end{tikzpicture}
    \caption{The considered scenario including a single-RX-RF HRIS, whose power splitting ratio and phase shifting network are controlled by a digital controller.}
    \label{fig:Scenario}
\end{figure}
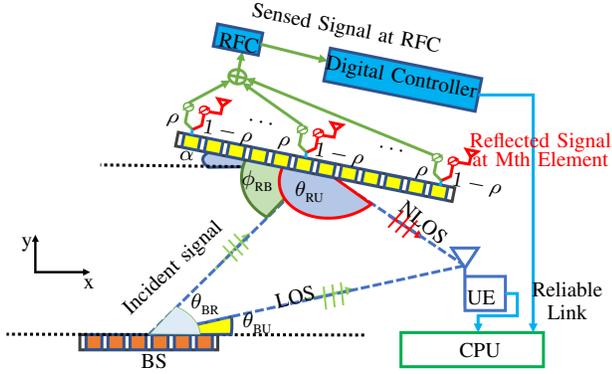

Consider the wireless system scenario in Fig.~\ref{fig:Scenario} operating in the mm-Wave frequency band and consisting of one $M_{_\text{B}}$-antenna BS with a known location $\mathbf{p}_\text{B}\in \mathbb{R}^2$, one single-antenna UE with an unknown location $\mathbf{p}_\text{U}\in \mathbb{R}^2$, and an HRIS with unknown location $\mathbf{p}_\text{R}\in \mathbb{R}^2$ and an orientation angle $\alpha$. We focus on the downlink direction considering that the BS sends $T$ orthogonal frequency division multiplexing (OFDM) symbols over time via $N_c$ subcarriers. We assume that each transmission time interval all the associated channels remain constant. We use the matrix $\mathbf{R}_{\alpha}$ to represent the rotation that maps the global frame of reference to the coordinate system associated with the HRIS. Moreover, the HRIS and UE are assumed not to be synchronized with the BS, leading to the unknown clock biases $b_\text{R}$ and $b_\text{U}$ at the HRIS and UE, respectively, with respect to the BS. Thus, besides the HRIS position and UE location, the clock bias of these components needs to be estimated.

We further assume that the UE operates in the FF range of the BS and the HRIS. The BS is equipped with a uniform linear array (ULA),
whose $m$-th element, with $m=0,1,\ldots,M_{_\text{B}}-1$, is located at the point $[\delta_\text{B}(2m-M_{_\text{B}}+1)/2, 0]^\top$, where $\delta_\text{B}$ denotes the spacing between adjacent antenna elements. 
Accordingly, we define the global coordinate system to be aligned with local coordinate system at the BS. The HRIS is a ULA with $M_{_\text{R}}$ unit elements attached to a single RX RF chain \cite{alexandropoulos2020hardware,alexandropoulos2021hybrid}, enabling the reception of the impinging signal. The $m$-th element, with $m=0,1,\ldots,M_{_\text{R}}-1$ of HRIS is located at the point (in the HRIS local coordinate system) $[\delta_\text{R}(2m-M_{_\text{R}}+1)/2, 0]^\top$, where $\delta_\text{R}$ is the space between two adjacent meta-elements.
For the sake of simplicity, we assume that $\delta_\text{R}= \delta_\text{B} =\delta=\lambda/2$, with $\lambda$ being the wavelength of the central carrier frequency. We finally make the common assumption \cite{Tsinghua_RIS_Tutorial} that the HRIS and the UE share their observations with a central processing unit (CPU), being responsible to carry out the targeted joint estimation that will be presented in the sequel, via a reliable link.
\vspace{-3mm}

\subsection{Models for the Received Signals and Channels}\label{sec:sig_model}
As shown in Fig.~\ref{fig:Scenario}, the HRIS structure includes $M_{_\text{R}}$ identical power splitters \cite{alexandropoulos2021hybrid}, which divide the received signal power at each element into two parts: one for reflection and the other for sensing/reception. For the latter operation, to feed a portion of the impinging signal to the single RX RF chain, the HRIS applies\footnote{A dedicated phase-shifting network feeds the impinging signals at all the HRIS elements to the single RX RF chain. It is assumed that the central processing unit knows both the HRIS combining vector and phase profile.}, a combining vector modeled by $\mathbf{c}_{_t}\in \mathbb{C}^{M_{_\text{R}}\times 1}$ with $|[\mathbf{c}_{_t}]_i| = 1$ during each time interval $t$. The HRIS also changes its phase profile in discrete time slots, to reflect signals towards the UE. We use the vector $\boldsymbol{\gamma}_t\in \mathbb{C}^{M_{_\text{R}} \times 1}$, with $|[\boldsymbol{\gamma}_{_t}]_k|= 1$, to represent the HRIS reflection phase profile at each time $t$. For the multi-antenna BS, we adopt the discrete Fourier transform (DFT) codebook for its beamforming vector $\mathbf{f}_{_t} \in \mathbb{C}^{M_{_\text{B}} \times 1}$ at each time instant $t$, and consider unit-power symbols for transmission.  
Under the latter assumptions and after removing the cyclic prefix
and taking the fast Fourier transform (FFT), the received OFDM symbols during each $t$-th time interval, with $t = 1,2,\ldots,T$ , 
at the HRIS and the UE, which are respectively denoted as $\mathbf{y}_{_\text{R}}(t)\in \mathbb{C}^{N_c\times 1}$ and $\mathbf{y}_{_\text{U}}(t)\in \mathbb{C}^{N_c\times 1}$, can be expressed as follows\footnote{For the sake of simplicity, we ignore the effects of scatter points (SPs) on the received signals, since their channels are weak compared to the HRIS channel. Nevertheless, we study how they affect our proposed estimator in our simulations.}: 
\begin{align}
    \mathbf{y}_{_\text{R}}(t) &= g_{_\text{BR}} \sqrt{\rho P_t}   \mathbf{d}(\tau_{_\text{BR}} )\, \mathbf{c}_{_t}^\top \mathbf{a}_{_\text{R}}(\phi_{_\text{RB}})\mathbf{a}^\top_{_\text{B}}(\theta_{_\text{BR}})\mathbf{f}_{_t}+\mathbf{w}_{_\text{R}}(t),\label{eq:y_ris}\\ 
    \mathbf{y}_{_\text{U}} (t) &= \mathbf{y}_{_\text{BU}} (t)+ \mathbf{y}_{_\text{BRU}} (t)+\mathbf{w}_{_\text{U}} (t),\label{eq:y_rx}
\end{align}
where $\mathbf{y}_{_\text{BU}}\in \mathbb{C}^{N_c\times 1}$ and $\mathbf{y}_{_\text{BRU}}\in \mathbb{C}^{N_c\times 1}$ represent the received signals at the UE from the BS through the line of sight (LOS) and reflected paths, respectively, which are given by:
\begin{align}
    \mathbf{y}_{_\text{BU}} (t)\, &=g_{_\text{BU}} \sqrt{P_t}   \mathbf{d}(\,\tau_{_\text{BU}} )\,\mathbf{a}^\top_{_\text{B}}(\theta_{_\text{BU}})\,\mathbf{f}_{_t}\label{eq:y_txrx},\\
    \mathbf{y}_{_\text{BRU}}(t) &= g_{_\text{BRU}}\sqrt{(1-\rho) P_t} \mathbf{d}(\tau_{_\text{BRU}} )\,\mathbf{a}^\top_{_\text{B}}({\theta}_{_\text{RU}})\,\text{diag} (\boldsymbol{\gamma}_{_t} ) \nonumber \\& \hspace{0.4cm}\mathbf{a}_{_\text{R}}(\phi_{_\text{RB}})\,\mathbf{a}^\top_{_\text{B}}(\theta_{_\text{BR}})\mathbf{f}_{_t},\label{eq:y_txRISrx}
\end{align}
where $\rho$ is the common power splitting ratio at the power splitters, while $g_{_\text{BU}}$, $g_{_\text{BR}}$, and $g_{_\text{BRU}}$ denote the unknown complex gains of the BS-UE, BS-HRIS, and BS-HRIS-UE wireless links, respectively, which are modeled as $g_{_i}\triangleq|g_{_i}| e^{-\jmath \phi_{_i}} $ with $i\in \{\text{BR},\text{BU},\text{BRU}\}$, where $\phi_{_i} \sim \mathcal{U}[0,2\pi)$ and $|g_{_i}|$ follows the model described in \cite{ghazalian2022bi}. The delay steering vector $\mathbf{d}(\tau)$ in the latter expressions is defined as: 
\vspace{-3mm}
\begin{equation}\label{eq:delay_tau}
    \mathbf{d}(\tau) \triangleq \left[ 1,\,e^{-\jmath 2\pi \Delta f \tau},\, \dots, \,e^{-\jmath 2\pi (N_c-1)\,\Delta f \tau} \right]^\top,
\end{equation}
where $\Delta f$ denotes the sub-carrier spacing. Based on the lack-of-synchronization assumption and the location of the different components of the system, we can express the propagation delay for all links as follows:
\begin{equation}\label{eq:tau}
    \tau_{_\text{BR}} = \frac{{d}_{_\text{BR}}}{c}+ b_{_\text{R}}, \,\tau_{_\text{BU}} = \frac{{d}_{_\text{BU}}}{c}+ b_{_\text{U}},  \, \tau_{_\text{BRU}} = \frac{{d}_{_\text{BR}}+{d}_{_\text{RU}}}{c}+ b_{_\text{U}}, 
\end{equation}
where ${d}_{_\text{BR}}\triangleq\Vert\mathbf{p}_{\text{B}}-\mathbf{p}_{\text{R}}\Vert$, ${d}_{_\text{BU}}\triangleq\Vert\mathbf{p}_{\text{B}}-\mathbf{p}_{\text{U}}\Vert$, and ${d}_{_\text{RU}}\triangleq\Vert\mathbf{p}_{\text{R}}-\mathbf{p}_{\text{U}}\Vert$ with $c$ being the speed of light.

In addition, $\theta_{_\text{BR}}$ and $\theta_{_\text{BU}}$ respectively represent the AODs from the BS towards the HRIS and the UE, based on the BS local coordinate system. In fact, the angles $\theta_{_\text{BR}}$ and $\theta_{_\text{BU}}$ are along the directions of the vectors $\mathbf{q}_{_\text{BR}}\triangleq (\mathbf{p}_{_\text{R}}-\mathbf{p}_{_\text{B}})/\Vert\mathbf{p}_{_\text{R}}-\mathbf{p}_{_\text{B}}\Vert$ and $\mathbf{q}_{_\text{BU}}\triangleq (\mathbf{p}_{_\text{U}}-\mathbf{p}_{_\text{B}})/\Vert\mathbf{p}_{_\text{U}}-\mathbf{p}_{_\text{B}}\Vert$, respectively.  Similarly, $\theta_{_\text{RU}}$ and $\phi_{_\text{RB}}$ are the AOD from the HRIS to the UE and the AOA to the HRIS from the BS both in the HRIS local coordinate system, respectively. Accordingly, one can write $\theta_{_\text{RU}} = \text{atan2}([\mathbf{q}_{_\text{RU}}]_2,[\mathbf{q}_{_\text{RU}}]_1)$ and $\phi_{_\text{RB}} = \text{atan2}([\mathbf{q}_{_\text{RB}}]_2,[\mathbf{q}_{_\text{RB}}]_1)$, where  $\mathbf{q}_{_\text{RU}}\triangleq\mathbf{R}_{\alpha}(\mathbf{p}_{_\text{U}}-\mathbf{p}_{_\text{R}})/\Vert\mathbf{p}_{_\text{U}}-\mathbf{p}_{_\text{R}}\Vert$ and $\mathbf{q}_{_\text{RB}}\triangleq \mathbf{R}_{\alpha}(\mathbf{p}_{_\text{B}}-\mathbf{p}_{_\text{R}})/\Vert\mathbf{p}_{_\text{B}}-\mathbf{p}_{_\text{R}}\Vert$.
Finally, in \eqref{eq:y_ris} and \eqref{eq:y_rx}, the steering vectors at the BS ($\mathbf{a}_{_\text{B}}(\nu) \in \mathbb{C}^{M_{_\text{B}}\times 1}$) and the HRIS ($\mathbf{a}_{_\text{R}}(\nu)\in \mathbb{C}^{M_{_\text{R}}\times 1}$) are defined as follows:
\begin{subequations}\label{eq:steer}
\begin{equation}\label{eq:a_B}
    \mathbf{a}_{_\text{B}}(\nu) = \left[e^{\jmath \frac{\pi\sin{\nu} (M_{_\text{B}}-1)}{2}}, \dots, 1, \dots, e^{-\jmath \frac{\pi\sin{\nu} (M_{_\text{B}}-1)}{2} }\right]^\top, 
\end{equation}
\begin{equation}\label{eq:a_R}
    \mathbf{a}_{_\text{R}}(\nu) = \left[e^{\jmath \frac{\pi\sin{\nu} (M_{_\text{R}}-1)}{2}}, \dots, 1, \dots, e^{-\jmath \frac{\pi\sin{\nu} (M_{_\text{R}}-1)}{2} }\right]^\top,
\end{equation}
\end{subequations}
$P_t$ denotes the transmitted power, and $\mathbf{w}_{_\text{R}}\in \mathbb{C}^{N_c\times 1}$ and $\mathbf{w}_{_\text{U}}\in \mathbb{C}^{N_c\times 1}$ represents the additive thermal noise vectors at the HRIS and the UE, respectively, each containing zero-mean circularly-symmetric independent and identically distributed Gaussian elements with variance\footnote{ $N_0$ is noise power spectral density.} $\sigma^2 = N_0 N_{_\text{c}} \Delta f$.
\vspace{-3mm}
\section{Proposed Multi-Parameter Estimation}

In this paper, capitalizing on the received signal models in \eqref{eq:y_ris} and \eqref{eq:y_rx}, we focus on the estimation of the unknown channel parameters included in the vector $\boldsymbol{\zeta}\triangleq[\boldsymbol{\eta_{_\text{ch}}}^\top,\mathbf{g}_{_\text{BR}}^\top,\mathbf{g}_{_\text{BU}}^\top,\mathbf{g}_{_\text{BRU}}^\top]^\top$,
where $\boldsymbol{\eta_{_\text{ch}}} \triangleq [\boldsymbol{\tau}^\top,\boldsymbol{\theta}^\top,\phi_{_\text{RB}}]^\top$ with $\boldsymbol{\tau} \triangleq [\tau_{_\text{BR}},\tau_{_\text{BU}},\tau_{_\text{BRU}}]^\top$ and $\boldsymbol{\theta} \triangleq [\theta_{_\text{BR}},\theta_{_\text{BU}},\theta_{_\text{RU}}]^\top$, as well as the state parameters in the vector $\boldsymbol{\zeta}_s\triangleq[\mathbf{p}_{_\text{R}}^\top, \alpha, \mathbf{p}_{_\text{R}}^\top, b_\text{R}, b_\text{U}]^\top$. We next present a multi-stage estimation approach that is based on the relationship derived in Section~\ref{sec:sig_model} between the channel and state parameters 
%

\subsubsection{BS-HRIS Channel Estimation} We commence by estimating the TOA at the HRIS. To this end, we stack all $T$ observations via  \eqref{eq:y_ris} in the matrix $\mathbf{Y}^0_{_\text{R}} \in \mathbb{C}^{N_c \times T}$, yielding: \vspace{-2mm}
\begin{align}\label{eq:y_ris_slack_TOA}
   \mathbf{Y}^0_{_\text{R}} &= g_{_\text{BR}} \sqrt{\rho P_t}  \mathbf{d}(\tau_{_\text{BR}} )\mathbf{a}_{_\text{RB}}^\top  +\mathbf{W}^0_{_\text{R}}, 
\end{align}
where $\mathbf{a}_{_\text{RB}}\triangleq \boldsymbol{\Omega}^\top \text{vec} (\mathbf{a}_{_\text{R}}(\phi_{_\text{RB}})\mathbf{a}^\top_{_\text{B}}(\theta_{_\text{BR}}))$, $\boldsymbol{\Omega}\triangleq[\mathbf{f}_1^\top \otimes \mathbf{c}_1^\top,\,\ldots,\,\mathbf{f}_T^\top \otimes \mathbf{c}_T^\top]^\top \in \mathbb{C}^{M_{_\text{B}}M_{_\text{R}}\times T}$, and $\mathbf{W}^0_{_\text{R}}\in \mathbb{C}^{N_c \times T}$ includes the noise contribution at the HRIS over all sub-carrier and time slots. We then compute the FFT of each column of $\mathbf{Y}^0_{_\text{R}}$ followed by integration over time (i.e., taking absolute value squared of the FFT output) to estimate $\tau _{_\text{BR}}$. For this, we use the method proposed in \cite{keykhosravi2021siso}. Then, we eliminate the effect of $\tau_{_\text{BR}}$ performing the calculation $\mathbf{Y}^0_{_\text{R}} \odot (\mathbf{d(-\hat{\tau}_{_\text{BR}}}) \mathbf{1}_{_\text{T}}^T)$. Taking the sum over the subcarriers and after some algebraic manipulations, the following expression is deduced:
\begin{align}\label{eq:y_ris0_AOA}
    \mathbf{y}^1_{_\text{R}}  &= g_{_\text{BR}} \sqrt{\rho P_t}   N_c \boldsymbol{\Omega}^\top \mathbf{A_{_\text{R}}}(\phi_{_\text{RB}}) \mathbf{a}_{_\text{B}}(\theta_{_\text{BR}})+\mathbf{w}^1_{_\text{R}}, 
\end{align}
where $\mathbf{y}^1_{_\text{R}} \in \mathbb{C}^{T \times 1}$,   $\mathbf{w}^1_{_\text{R}}\triangleq (\mathbf{W}^0_{_\text{R}})^\top \odot (\mathbf{1}_{_\text{T}}\mathbf{d}(-\hat{\tau}_{_\text{BR}})^\top) \mathbf{1}_{_{\text{N}_c}}$, and $\mathbf{A_{_\text{R}}}(\phi_{_\text{RB}})\triangleq (\,\mathbf{I}_{_{M_{_\text{R}}}}\otimes \mathbf{a}_{_\text{R}}(\phi_{_\text{RB}}))$. Using this expression, we can obtain the estimates for $g_{_\text{BR}}$, $\theta_{_\text{BR}}$, and $\phi_{_\text{RB}}$ via the ML problem:
\begin{align}\label{eq:MLE_HRIS_BS link}
  \left[\hat{g}_{_\text{BR}},\hat{\theta}_{_\text{BR}},\hat{\phi}_{_\text{RB}} \right] =  \arg\min_{g_{_\text{BR}},\theta_{_\text{BR}},\phi_{_\text{RB}}} \Vert&\mathbf{y}^1_{_\text{R}} -g_{_\text{BR}} \sqrt{\rho P_t}   N_c \boldsymbol{\Omega}^\top \nonumber\\&\mathbf{A_{_\text{R}}}(\phi_{_\text{RB}})\, \mathbf{a}_{_\text{B}}(\theta_{_\text{BR}})\Vert^2.
\end{align}

To solve the latter optimization problem, we introduce the unstructured vector $\mathbf{v}\triangleq g_{_\text{BR}} \sqrt{\rho P_t}N_c \mathbf{a}_{_\text{B}}(\theta_{_\text{BR}})$. We first estimate $\phi_{_\text{RB}}$ via the following minimization problem:
\begin{align}\label{eq:MLE_phi_br}
  \hat{\phi}_{_\text{RB}} &=  \arg\min_{\phi{_\text{RB}}} \Vert\mathbf{y}^1_{_\text{R}} -\boldsymbol{\Omega}^\top\mathbf{A_{_\text{R}}}(\phi_{_\text{RB}})\mathbf{v} (\phi_{_\text{RB}})\Vert^2,
\end{align}
where $\mathbf{v} (\phi_{_\text{RB}}) = (\boldsymbol{\Omega}^\top \mathbf{A_{_\text{R}}}(\phi_{_\text{RB}}))^\dag\mathbf{y}^1_{_\text{R}}$ represents the estimate of $\mathbf{v}$ as a function of $\phi_{_\text{RB}}$. The solution of \eqref{eq:MLE_phi_br} can be obtained through a line search over $\phi_{_\text{RB}}$.
Considering \eqref{eq:y_ris0_AOA} and \eqref{eq:MLE_phi_br}, we can also estimate $\theta_{_\text{BR}}$ via the minimization:
\begin{align}\label{eq:MLE_theta_br}
  \hat{\theta}_{_\text{BR}} &=  \arg\max_{\theta_{_\text{BR}}} \Vert\mathbf{y}^1_{_\text{R}} -g_{_\text{BR}}(\theta_{_\text{BR}}) \beta \boldsymbol{\Omega}^\top \mathbf{A_{_\text{R}}}(\hat{\phi}_{_\text{RB}}) \mathbf{a}_{_\text{B}}(\theta_{_\text{BR}})\Vert^2,
\end{align}
where $\beta\triangleq\sqrt{\rho P_t} N_c$ and $g_{_\text{BR}}$ can be expressed as a function of $\theta_{_\text{BR}}$ as $g_{_\text{BR}}(\theta_{_\text{BR}}) = (\beta \boldsymbol{\Omega}^\top \mathbf{A_{_\text{R}}}(\hat{\phi}_{_\text{RB}}) \mathbf{a}_{_\text{B}}(\theta_{_\text{BR}}))^\dag \mathbf{y}^1_{_\text{R}}$. 
Note that, one can solve \eqref{eq:MLE_theta_br} via a line search over $\theta_{_\text{BR}}$. All in all, we use Newton's method to refine the estimation of $\theta_{_\text{BR}}$ and $\phi_{_\text{RB}}$ considering \eqref{eq:MLE_HRIS_BS link}, while applying $\hat{\phi}_{_\text{RB}}$ and $\hat{\theta}_{_\text{BR}}$ obtained in \eqref{eq:MLE_phi_br} and \eqref{eq:MLE_theta_br}, respectively, as the method's initial points.
\subsubsection{Estimation of the BS-UE and BS-HRIS-UE Channels}
We assume that the received power at the UE from the BS is much larger than that from the HRIS, which facilitates us to first estimate the BS-UE channel
\footnote{For scenarios where the HRIS has large dimensions, a certain condition in the time evolution of the phase profile has been imposed [15]. This can be adopted to cancel the effect of the HRIS channel, and hence, assist in the channel estimation process.}. 
Based on this assumption as well as on \eqref{eq:y_rx} and \eqref{eq:y_txrx}, we can approximate the received signal at the UE as follows: 
\begin{equation}\label{eq:yrx_approxi}
    \mathbf{Y}^0_{_\text{U}} \approx g_{_\text{BU}} \sqrt{P_t}   \mathbf{d}(\tau_{_\text{BU}} )\,\mathbf{a}^\top_{_\text{B}}(\theta_{_\text{BU}})\mathbf{F} + \mathbf{W}^0_{_\text{BU}},
\end{equation}
where $\mathbf{F} \triangleq[\mathbf{f}_1,\,\mathbf{f}_2,\,\dots,\,\mathbf{f}_T]$ and $\mathbf{W}^0_{_\text{BU}} \in \mathbf{R}^{N_{_\text{c}}\times T}$
includes the noise contribution at the UE over all subcarriers and
time slots. We take a similar approach to that before to estimate $\tau_{_\text{BU}}$. After removing the effect of this TOA and integrating the signals over the $N_c$ subcarrier frequencies, the following expression is deduced:
\begin{equation}\label{eq:y_txue_integ}
     \mathbf{y}^0_{_\text{U}}= g_{_\text{BU}} \sqrt{P_t} N_c \mathbf{F}^\top \mathbf{a}_{_\text{B}}(\theta_{_\text{BU}})+ \mathbf{w}^0_{_\text{BU}}, 
\end{equation}
where $\mathbf{y}^0_{_\text{U}} \in \mathbb{C}^{T\times 1}$ and $\mathbf{w}^0_{_\text{BU}}\triangleq  (\mathbf{W}^0_{_\text{BU}})^\top \odot (\mathbf{1}_{_\text{T}}\mathbf{d}(-\tau_{_\text{BU}})^\top) \mathbf{1}_{_{N_c}}$. The parameter $\theta_{_\text{BU}}$ can be estimated by solving the following problem via a simple line search:
\begin{align}\label{eq:MLE_theta_bU}
  \hat{\theta}_{_\text{BU}} &=  \arg\min_{\theta_{_\text{BU}}} \Vert\mathbf{y}^0_{_\text{U}} -g_{_\text{BU}}(\,\theta_{_\text{BU}})\, \sqrt{P_t} N_c \mathbf{F}^\top \mathbf{a}_{_\text{B}}(\,\theta_{_\text{BU}})\, \Vert^2,
\end{align}
where $g_{_\text{BU}}(\theta_{_\text{BU}})\triangleq (\sqrt{P_t} N_c \mathbf{F}^\top \mathbf{a}_{_\text{B}}(\theta_{_\text{BU}}) )^\dag\mathbf{y}^0_{_\text{U}}$. This coarse estimate can be used as the initial point in the Newton's method, as described before.

By using the BS-UE channel estimation, we can recover the LOS signal received at the UE from the BS, and consequently, remove its contribution from the received signal. However, it is critical to note that one needs to estimate the LOS channel parameters with high accuracy, otherwise, they will affect the estimation of the NLOS channel parameters. We proceed by defining the matrix:
\begin{equation}\label{eq:y_reflect_compact}
 \mathbf{Y}^1_{_\text{U}} \approx g_{_\text{BRU}}\sqrt{(\,1-\rho)\, P_t} \mathbf{d}(\,\tau_{_\text{BRU}} )\,\mathbf{a}^\top (\,\theta_{_\text{RU}},\hat{\theta}_{_\text{BR}},\hat{\phi}_{_\text{RB}})\, \boldsymbol{\Xi}+ \mathbf{W}^0_{_\text{BRU}},
\end{equation}
where $\boldsymbol{\Xi}\triangleq \mathbf{F}\otimes\boldsymbol{\Gamma}^\top$, $\boldsymbol{\Gamma} \triangleq [\boldsymbol{\gamma}_{1}, \dots, \boldsymbol{\gamma}_{T}]$,  $\mathbf{W}^0_{_\text{BRU}}\in \mathbb{C}^{N_c \times T}$, and 
\begin{equation}\label{eq:a}
\mathbf{a} (\theta_{_\text{RU}},\hat{\theta}_{_\text{BR}},\hat{\phi}_{_\text{BR}})\triangleq \text{vec}((\mathbf{a}_{_\text{R}}(\theta_{_\text{RU}})\, \odot \mathbf{a}_{_\text{R}}(\hat{\phi}_{_\text{RB}})) \mathbf{a}^\top_{_\text{B}}(\hat{\theta}_{_\text{BR}})).
\end{equation}
It is noted that the adoption of the HRIS allows us to estimate the AOA and AOD at the HRIS side, which would be impossible in the case of a passive RIS.
Similarly to above, after removing the effect of $\tau_{_\text{BRU}}$ from \eqref{eq:y_reflect_compact} and integrating over all sub-carrier frequencies, we can obtain the expression:
\begin{equation}\label{eq:y_r_bru_remov}
 \mathbf{y}^1_{_\text{U}} = \tilde{g} \boldsymbol{\Xi}^\top\mathbf{a} (\,\theta_{_\text{RU}},\hat{\theta}_{_\text{BR}},\hat{\phi}_{_\text{RB}})\, + \mathbf{w}^0_{_\text{BRU}}, 
\end{equation}
where $\tilde{g}\triangleq g_{_\text{BRU}}\sqrt{(1-\rho) P_t}N_c$ and $\mathbf{w}^0_{_\text{BRU}}\triangleq  (\mathbf{W}^0_{_\text{BRU}})^\top \odot (\mathbf{1}_{_\text{T}}\mathbf{d}(-\tau_{_\text{BRU}})^\top) \mathbf{1}_{_{N_c}}$. Hence, the estimation for angle $\theta_{_\text{RU}}$ can be obtained from the following minimization:
\begin{align}\label{eq:MLE_theta_RU}
  \hat{\theta}_{_\text{RU}} &=  \arg\min_{\theta_{_\text{RU}}} \Vert\mathbf{y}^1_{_\text{U}} - \tilde{g}_{_\text{BRU}}(\theta_{_\text{RU}}) \boldsymbol{\Xi}^\top \mathbf{a} (\theta_{_\text{RU}},\hat{\theta}_{_\text{BR}},\hat{\phi}_{_\text{RB}}) \Vert^2,
\end{align}
with $\tilde{g}_{_\text{BRU}}(\,\theta_{_\text{RU}})\triangleq (\boldsymbol{\Xi}^\top \mathbf{a}(\,\theta_{_\text{RU}},\hat{\theta}_{_\text{BR}},\hat{\phi}_{_\text{RB}})\,)^\dag\mathbf{y}^1_{_\text{U}}$, using a simple line search. As before, this coarse estimation of $\hat{\theta}_{_\text{RU}}$ can be used as the initial point of the Newton's method for solving the ML optimization problem \eqref{eq:MLE_theta_RU}.


\begin{figure*}
\centering
\setlength{\tabcolsep}{0pt}
\begin{tabular}{c c}
\subfloat[]{\includegraphics[width=.24\linewidth]{Figure/PEB_pt.tikz}
\label{fig: PEB_pt}}
\subfloat[]{\includegraphics[width=.24\linewidth]{Figure/TEB_pt.tikz}
\label{fig: TEB_pt}}
\subfloat[]{\includegraphics[width=.24\linewidth]{Figure/Angles_pt.tikz}
\label{fig: angle_pt}}
\subfloat[]{\includegraphics[width=.24\linewidth]{Figure/TEB_PEB_rho.tikz}
\label{fig: PEB_TEB_rho}}\\
\subfloat[]{\includegraphics[width=.24\linewidth]{Figure/Angles_rho.tikz}
\label{fig: angles_rho}}
\subfloat[]{\includegraphics[width=.28\linewidth]{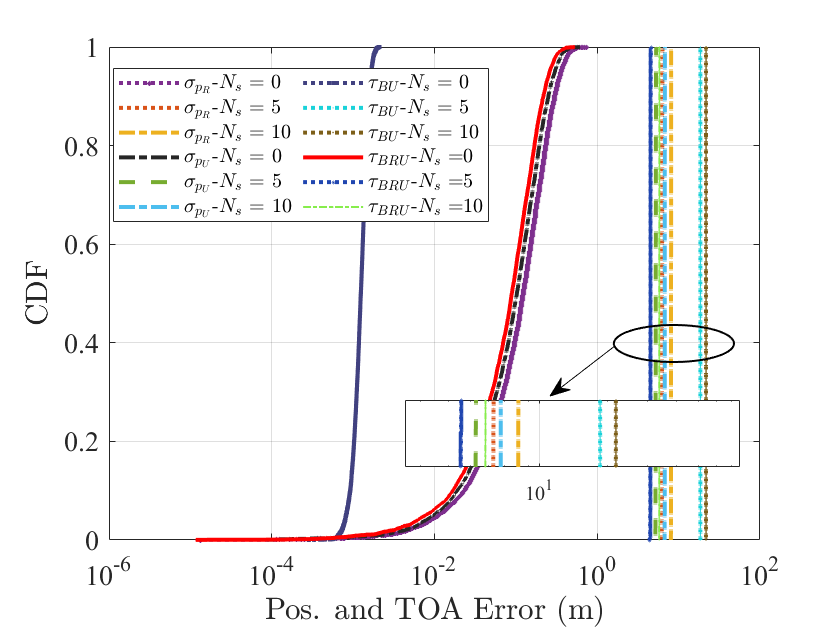}
\label{fig: PEB_cdf}}
\subfloat[]{\includegraphics[width=.28\linewidth]{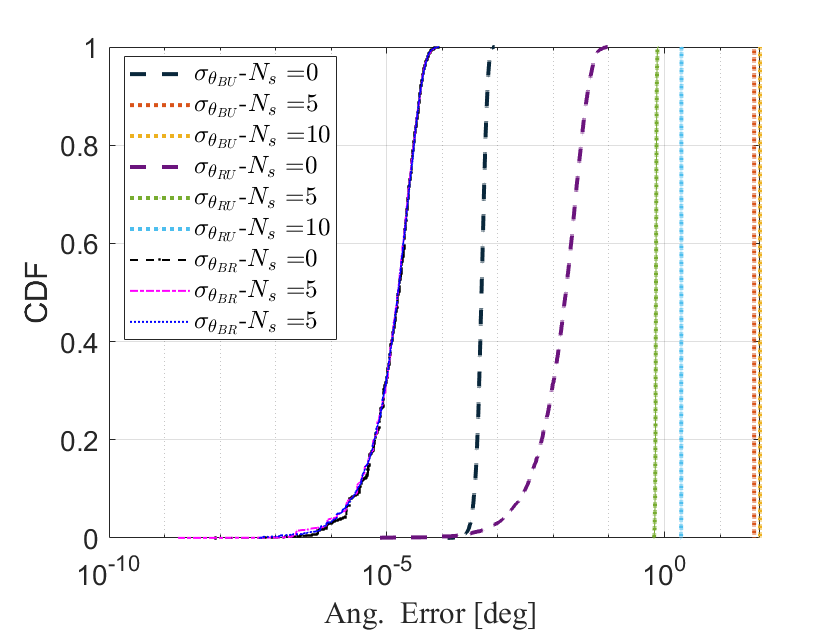}
\label{fig: angles_cdf}}
\end{tabular}
\caption{
The evaluation of the proposed estimator for the case where SPs are randomly distributed in $[x,5]$, with $x\sim \mathcal{U}(8,13)$, their radar cross section is $0.1m^2$, and the power splitting factor is set to $\rho=0.5$. (a) The RMSE of the position estimations; (b) The RMSE of the ToAs estimations; (c) The RMSE of the AOA/AODs and RIS orientation estimations; (d) The effect of $\rho$ on the TOA/positions estimations; (e) The effect of $\rho$ on the angles (i.e., AoAs/AoDs and $\alpha$) estimations; (f) The effect of the number of SPs around the UE on the estimation of the TOA and positions; and (g) The effect of the number of SPs around the UE on the estimation of the angles.}
\label{fig:che}
\end{figure*}

\subsubsection{State Estimation}
In this step, the state parameters are estimated using the previously estimated parameters. We make use of the one-to-one mapping presented in Section~\ref{sec:sig_model} between the channel and state parameters. We specifically use expression \eqref{eq:tau} to define the following parameter:\vspace{-1mm}
\begin{equation}\label{eq:TDOA}
   \hat{{d}}\triangleq{d}_{_\text{BR}}+{d}_{_\text{RU}}-{d}_{_\text{BU}}= c (\hat{\tau}_{_\text{BRU}}-\hat{\tau}_{_\text{BU}}).
\end{equation}
\vspace{0mm}
Based on the triangular shape formulated by the system nodes in Fig.~\ref{fig:Scenario}, we can write ${\sin{(\beta_0)}}/{{d}_{_\text{RU}}} = {\sin{(\beta_1)}}/{{d}_{_\text{BU}}}  = {\sin{(\beta_2)}}/{{d}_{_\text{BR}}}, $
where $\angle\beta_0 = |\hat{\theta}_{_\text{BR}}-\hat{\theta}_{_\text{BU}}|$, $\angle\beta_1 = |\hat{\phi}_{_\text{RB}}-\hat{\theta}_{_\text{RU}}|$, and $\angle\beta_2 = \pi -\angle\beta_0 - \angle\beta_1$. From \eqref{eq:TDOA} and considering the law of sines, the distances from the BS to the other two network nodes can be computed as:
\vspace{-0mm}
\begin{equation}\label{eq:dtx-rx}
\hat{d}_{_\text{BU}} = \frac{\hat{d} \sin{\beta_1}}{\sin{\beta_0}+ \sin{\beta_2}-1}, \,\hat{d}_{_\text{BR}} = \frac{\hat{d} \sin{\beta_2}}{\sin{\beta_0}+ \sin{\beta_2}-1}.
\end{equation}
Using these expressions and the AODs from the BS to the other two nodes, the positions of the HRIS and UE can be estimated as $\hat{\mathbf{p}}_{_\text{R}}=\mathbf{p}_{_\text{B}}+\hat{d}_{_\text{BR}}\cos{\hat{\theta}_{_\text{BR}}}$ and $\hat{\mathbf{p}}_{_\text{U}}=\mathbf{p}_{_\text{B}}+\hat{d}_{_\text{BU}}\cos{\hat{\theta}_{_\text{BU}}}$, respectively. We can also estimate the clock bias at the HRIS and UE as $\hat{b}_{_\text{R}} =\hat{\tau}_{_\text{BR}} - \hat{d}_{_\text{BR}}/c$ and $\hat{b}_{_\text{U}} =\hat{\tau}_{_\text{BU}} - \hat{d}_{_\text{BU}}/c$, respectively. Based on the relationship between the AOA at the HRIS from the BS and the AOD from the BS to the HRIS, one can finally estimate the HRIS orientation as $\hat{\alpha} = \pi - \hat{\theta}_{_\text{BR}}-\hat{\phi}_{_\text{RB}}$. 

\section{Numerical Results}
In this section, we compare the root mean square error (RMSE)\footnote{We use the notations $\sigma_{_\text{R}}$ and $\sigma_{_\text{U}}$ for the RMSE of the HRIS position estimate $\hat{\mathbf{p}}_{_\text{R}}$ and the UE position $\hat{\mathbf{p}}_{_\text{U}}$, respectively. The RMSE of the HRIS orientation estimation $\hat{\alpha}$ is denoted by $\sigma_{_\alpha}$, while $\sigma_{\tau_\text{BR}}$, $\sigma_{\tau_\text{BU}}$, and $\sigma_{\tau_\text{BRU}}$ are the RMSE of estimations of the parameters $\hat{\tau}_{_\text{BR}}$, $\hat{\tau}_{_\text{BU}}$, and $\hat{\tau}_{_\text{BRU}}$, respectively. Besides, The RMSE of the angle estimations $\hat{\theta}_{_\text{BR}}$, $\hat{\theta}_{_\text{BU}}$, $\hat{\theta}_{_\text{RU}}$, and $\hat{\phi}_{_\text{RB}}$ are represented by $\sigma_{\theta_{\text{BR}}}$, $\sigma_{\theta_{\text{BU}}}$, $\sigma_{\theta_{\text{RU}}}$, and $\sigma_{\phi_{\text{RB}}}$, respectively.} of the estimated parameters with the corresponding CRBs\footnote{The CRBs corresponding to the estimations of the AoAs, AoDs, and ToAs, position and orientation are henceforth termed as AOA error bound (AAEB), AOD error bound (ADEB), TOA error bound (TEB), position error bound (PEB), and orientation error bound (OEB), respectively.} to evaluate the proposed estimator. For the RMSE calculations, we have averaged the results over $500$ independent noise realizations. All the phase shifts of the HRIS elements have been drawn from the uniform distribution, i.e., $\angle[\boldsymbol{\gamma}_{_t}]_{k} \sim \mathcal{U}[0,2\pi)$. The same holds for the combiner vector at the HRIS, i.e., $\angle[\mathbf{c}_{_t}]_{i} \sim \mathcal{U}[0,2\pi)$. The rest of the simulation parameters are given in Table~\ref{table: tab1}. 
 \begin{table}[t!]
\caption{The considered simulation parameters.}
\begin{center}
\begin{tabular}{l c c} 
 \hline \hline
 Parameter & Symbol & Value  \\  
 \hline\hline
 Wavelength & $\lambda$ & $1 ~\text{cm}$
\\Light speed & $c$ & $3\times 10^{8}~ \text{m}/\text{sec}$\\
 Number of subcarriers & $N_c$& $100$\\
 Number of transmissions & $T$& $32$\\
 Sub-carrier spacing & $\Delta f$& $120~\text{kHz}$\\
 Noise PSD&$N_0$& $-174~\text{dBm/Hz}$\\ RX's noise figure& $n_f$ & $5~\text{dB}$\\ IFFT Size & $N_F$& $1024$\\ UE position& $\mathbf{p}_\text{U}$& $\left[6\text{m},6\text{m}\right]^\top$  \\BS position& $\mathbf{p}_{\text{B}}$& $\left[0\text{m},0\text{m}\right]^\top$ \\HRIS position& $\mathbf{p}_{\text{R}}$& $\left[2\text{m},10\text{m}\right]^\top$\\
 RIS orientation& $\alpha$& $\pi/6~\text{rad}$
 \\Number of BS antennas& $M_{_\text{B}}$& 17\\Number of HRIS elements& $M_{_\text{R}}$& 33\\
 \hline\hline
\end{tabular}
\label{table: tab1}
\end{center}
\end{table}

We first study the effect of the transmitted power $P_t$ on the RMSE of the estimated parameters, as shown in Figs.~\ref{fig: PEB_pt}--~\ref{fig: angle_pt}. As can be seen, the RMSE of the estimations for the channel and the state parameters decreases as $P_t$ increases, and the bounds for the channels parameters' estimations (i.e., BS-UE, BS-HRIS, BS-HRIS-UE) are attained when $P_t = 0$ dBm and beyond. However, the RMSE of the BS-UE channel parameters' estimation deviate from its corresponding CRB when $P_t >5$ dBm. This happens because the error caused by noise is much smaller than that caused by the considered approximation at large $P_t$ values, i.e.,  expression~ \eqref{eq:yrx_approxi} becomes inaccurate. This implies that the larger $P_t$ is, the higher is the level of the reflected signals from the HRIS towards the UE. 

In Figs.~\ref{fig: PEB_TEB_rho}--~\ref{fig: angles_rho}, the effect of the common power splitting factor $\rho$ on the estimation performance is assessed. As can be observed, as $\rho$ increases (i.e., the sensing/reception power at the HRIS increases), the CRB of the BS-HRIS channel parameters' estimation decreases. We can see that the CRB of the HRIS rotation angle also decreases. This is due to the fact that the estimation of $\alpha$ directly depends on the estimation of BS-HRIS channel parameters (i.e., $\hat{\theta}_{_\text{BR}}$ and $\hat{\phi}_{_\text{RB}}$). Moreover, it is demonstrated that, as $\rho$ increases, the estimation of the position of the HRIS and the UE exhibits a higher error variance, since these positions' estimations are coupled. Increasing the value of $\rho$ leads to a decrease in the signal-to-noise ratio (SNR) at the UE. Accordingly, the UE cannot estimate the reflected channels, which contain jointly information about the UE and the HRIS, with high accuracy. We also observe that when $\rho\to 0$ and $\rho\to 1$, the estimations of both the BS-HRIS and HRIS-UE channel parameters fail. It can be also seen that $\rho$ has no effect on the BS-UE channel estimation.  

Finally, we depict the effect of SPs around HRIS on the proposed estimator accuracy in Figs.~\ref{fig: PEB_cdf}--~\ref{fig: angles_cdf}. Since similar algorithms are used for both BS-RIS and BS-UE channels, we only consider the SPs effect on latter channel, when evaluating their effect on estimation accuracy using the approach of \cite{keykhosravi2021semi}. It can be seen that, as the number $N_{_\text{s}}$ of SPs increases, the accuracy of the estimations deteriorates. These deterioration are mostly on the $\tau_{_\text{BU}}$ and $\theta_{_\text{BU}}$. To tackle this issue, one solution can be the adoption of orthogonal profiles at the HRIS over time \cite{keykhosravi2022ris}; we leave this approach for future work. 

\section{Conclusion}
In this paper, we presented a multi-stage estimator for the unknown location and orientation of a single-RX-RFC HRIS and the unknown position of a single-antenna UE in a multi-carrier system with a multi-antenna BS. The proposed estimation leverages the channel's geometrical features 
to estimate the targeted parameters. We showed that the RMSE of the estimations approaches the corresponding CRBs within a certain SNR range, and identified the critical role of the HRIS power splitting ratio on the estimation accuracy. 
\balance 
\bibliographystyle{IEEEtran}
\bibliography{ref.bib}
\end{document}